\pdfoutput=1
\newlength{\absize}
\documentclass[12pt]{article}
\usepackage[T1]{fontenc}
\usepackage{graphicx,dsfont,subfig,amsmath,amssymb,setspace,sectsty,float,color,cite,slashed}
\usepackage[ocgcolorlinks]{hyperref}
\hypersetup{linktocpage}
\hypersetup{linkcolor=blue,citecolor=blue,urlcolor=blue}
\renewcommand{\baselinestretch}{1.5}

\sectionfont{\Large}
\numberwithin{equation}{section}
\DeclareMathOperator{\sech}{sech}

\begin{document}
\thispagestyle{empty}
\pagestyle{empty}
\renewcommand{\thefootnote}{\fnsymbol{footnote}}
\newcommand{\starttext}{\newpage\normalsize
\pagestyle{plain}
\setlength{\baselineskip}{3.5ex}\par
\setcounter{footnote}{0}
\renewcommand{\thefootnote}{\arabic{footnote}}}
\newcommand{\preprint}[1]{\begin{flushright}
\setlength{\baselineskip}{3ex}#1\end{flushright}}
\renewcommand{\title}[1]{\begin{center}\Large\bf
#1\end{center}\par}
\renewcommand{\author}[1]{\vspace{2ex}{\normalsize\begin{center}
\setlength{\baselineskip}{3.25ex}#1\par\end{center}}}
\renewcommand{\thanks}[1]{\footnote{#1}}
\renewcommand{\abstract}[1]{\vspace{2ex}\small\begin{center}
\centerline{\bf Abstract}\par\vspace{2ex}\parbox{\absize}{#1
\setlength{\baselineskip}{3.25ex}\par}
\end{center}}
\setcounter{bottomnumber}{2}
\setcounter{topnumber}{3}
\setcounter{totalnumber}{4}
\renewcommand{\bottomfraction}{1}
\renewcommand{\topfraction}{1}
\renewcommand{\textfraction}{0}
\def\draft{
\renewcommand{\label}[1]{{\quad[\sf ##1]}}
\renewcommand{\ref}[1]{{[\sf ##1]}}
\renewenvironment{equation}{$$}{$$}
\renewenvironment{thebibliography}{\section*{References}}{}
\renewcommand{\cite}[1]{{\sf[##1]}}
\renewcommand{\bibitem}[1]{\par\noindent{\sf[##1]}}
}
\def\theequation{\thesection.\arabic{equation}}
\preprint{}
\newcommand{\be}{\begin{equation}}
\newcommand{\ee}{\end{equation}}
\newcommand{\ba}{\begin{eqnarray}}
\newcommand{\ea}{\end{eqnarray}}
\newcommand{\bas}{\begin{eqnarray*}}
\newcommand{\eas}{\end{eqnarray*}}
\newcommand{\bc}{\begin{center}}
\newcommand{\ec}{\end{center}}
\newcommand{\nn}{\nonumber}
\newcommand{\comment}[1]{}
\vspace{2ex}
\title{Killing Spinor in Linearly Confining Supergravity}
\author{Girma Hailu\thanks{{\color{blue}hailu@physics.harvard.edu}}\\
\vspace{2ex}
\it{Department of Physics\\
Harvard University\\
Cambridge, MA 02138} }

\abstract{The killing spinor of a linearly confining supergravity background previously proposed and argued to produce features of pure $\mathcal{N}=1$ $SU(N)$ gauge theory in four dimensions is constructed directly using the supersymmetry variations of the gravitino and the dilatino.
}

\starttext
\newpage

\section{\label{sec:intro}Introduction}

We previously proposed and argued in \cite{Hailu:2011pn} a supergravity theory producing features of pure $\mathcal{N}=1$ $SU(N)$ gauge theory in four dimensions. It was shown in \cite{Hailu:2011kp} that the background produces linear confinement of quarks in four dimensions. In \cite{Hailu:2012jc}, mass spectrum of $0^{++}$ glueballs agreeing with available lattice data from large $N$ non-supersymmetric lattice QCD was produced. A cosmological model was constructed in \cite{Hailu:2012ut} in which an accelerated expansion followed by a smooth transition to decelerating expansion is dynamically generated. More detail about linear confinement was also given in \cite{Hailu:2012ut} where the strong interaction was used to argue for stabilization of the model universe during Friedmann evolution, and dark energy arising from the interaction of the model universe with the background. Because the supergravity theory provides a realization of all these features on the same setting and the solutions are given by analytic expressions, it is important to further investigate the background and possibilities for building additional models. The supergravity solutions were obtained using the equations of motion in \cite{Hailu:2007ae, Grana:2004bg, Butti:2004pk} which were written using $SU(3)$ structures to organize the balance between flux and torsion.

Let us summarize the type IIB supergravity solutions for the metric and the fluxes in \cite{Hailu:2011pn}. The 10D metric is given by
\ba
&{d{s}_{10}}^2=\cosh u\,dx_{1,3}^2
+ r_s^2\,\sech u\,(d\rho^2+d\psi^2&\nn\\&+d\varphi_1^2+d\varphi_2^2
+d\varphi_3^2+d\varphi_4^2),\label{metric-3n}&
\ea
where
\be
\tanh u= \frac{g_sN}{2\pi}\rho,\quad \rho\equiv 3 \ln (\frac{r}{r_s}),\label{hn-1}
\ee
and $dx_{1,3}^2$ is the metric on flat four-dimensional spacetime $\mathds{R}^{1,3}$ with coordinate $x^\mu$, $g_s$ is the string coupling, and $N$ is number of $D7-$branes. The extra 6D space of the 10D spacetime is parameterized by one radial coordinate $r$ with range $r_s\le r\le r_s\,e^{\frac{2\pi}{3g_sN}}$ and five angles $\psi$, $\varphi_1$, $\varphi_2$, $\varphi_3$, and $\varphi_4$, each having the range between 0 and $2\pi$.  Upper case indices $M,\,N,\,\cdots$ are used to represent the coordinates of the 10D spacetime, Greek letters $\mu,\,\nu,\,\cdots$ label the coordinates of the 4D spacetime, and lower case indices  $m,\,n,\,\cdots$ denote the coordinates of the extra 6D space.
The fluxes and the dilaton are given by
\ba
 &F_1=\frac{N}{2\pi}d\psi,&\nn\\ &F_3=\frac{N}{2\pi}\,r_s^2\,\tanh u \, \left(d\psi\wedge d\varphi_1\wedge  d\varphi_2 +d\psi\wedge  d\varphi_3\wedge  d\varphi_4\right),&\nn\\ &H_3=\frac{g_sN}{2\pi}\,r_s^2\,\left(d\rho\wedge d\varphi_1\wedge  d\varphi_2+d\rho\wedge d\varphi_3\wedge  d\varphi_4\right), &\nn\\
&\tilde{F}_5=(1+\star_{10})F_5=\frac{N}{2\pi}\,r_s^4\,\cosh 2u\, \sech^2u  \, d\psi\wedge d\varphi_1\wedge  d\varphi_2\wedge  d\varphi_3\wedge  d\varphi_4&\nn\\ &-\frac{N}{2\pi}\,\,\cosh 2u\, \cosh^2u \, d^4x \wedge d\rho,
\qquad \,& \nn\\
&e^{\Phi}=g_s.&
\label{F5sol-1bbn}
\ea
$F_1$, $F_3$, and $\tilde{F}_5$ are the 1-form, 3-form, and 5-form R-R fluxes.  $\tilde{F}_5$ is self-dual, $H_3$ is the NS-NS 3-form flux, and $\Phi$ is the dilaton.
The fluxes are all induced by $N$ D7-branes wrapping a 4-cycle with coordinates $\varphi_1$, $\varphi_2$, $\varphi_3$, and $\varphi_4$ at $r=r_s$.
It was shown in \cite{Hailu:2011pn} that the metric and fluxes above solve each and every one of the bosonic supergravity equations of type IIB theory.

In this note, we show that the above metric and fluxes solve the supersymmetry variations of the gravitino and the dilatino of type IIB theory directly and construct the corresponding killing spinor for $\mathcal{N}=1$ supersymmetry in four dimensions. Preserving supersymmetry is important in order to have a stable background geometry. Understanding structure of the killing spinor is useful for investigating supersymmetric and non-supersymmetric embeddings of $D-$branes.

\section{\label{sec:kspinor}Killing spinor}

It will be convenient to work in a basis with constant gamma matrices. Therefore, first we rewrite the metric in terms of coordinate one-forms as
\ba
{d{s}_{10}}^2=G_{\hat{M}\hat{N}}e^{\hat{M}}e^{\hat{N}},
\ea
where $G_{\hat{M}\hat{N}}=diag(-1,1,\cdots,1)$ with the hat in $\hat{M}$ and $\hat{N}$ used to distinguish the basis,
\be
e^{\hat{M}}=E^{\hat{M}}_{\,\,\,N}dx^{N},
\ee
and $E^{\hat{M}}_{\,\,\,N}$ is the vielbein. More explicitly,
\ba
&e^{\hat{\mu}}=\sqrt{\cosh u}\, dx^\mu,\quad e^{\hat{4}}=r_s\sqrt{\sech u}\, d\rho,\quad e^{\hat{5}}=r_s\sqrt{\sech u}\, d\psi,&\nn\\ &  e^{\hat{6}}=r_s\sqrt{\sech u}\,d\varphi_1, \quad e^{\hat{7}}=r_s\sqrt{\sech u}\, d\varphi_2,&\nn\\ &
e^{\hat{8}}=r_s\sqrt{\sech u}\, d\varphi_3,\quad e^{\hat{9}}=r_s\sqrt{\sech u}\, d\varphi_4,&
\ea
and
\be
E^{\hat{\mu}}_{\,\,\,N}=\sqrt{\cosh u} \,\delta^\mu_{\,\,\,N},\qquad E^{\hat{m}}_{\,\,\,N}=r_s\sqrt{\sech u} \, \delta^m_{\,\,\,N}.
\ee

The supersymmetry transformations of the gravitino and the dilatino fields in 10D can be expressed, see \cite{Bergshoeff:2001pv} for instance,  as
\be\label{st-psi}
\delta \psi_{\hat{M}} = \nabla_{\hat{M}}\,\epsilon -\frac{1}{4} (\slashed{H}_3)_{\hat{M}}\sigma^3\,\epsilon+\frac{1}{8}e^{\Phi}\Bigl( \slashed{F}_{3}\sigma^1
+i(\slashed{F}_{1}+\frac{1}{2}\slashed{\tilde{F}}_{5})\sigma^2\Bigr)\Gamma_{\hat{M}}\,\epsilon,
\ee
\be\label{st-lam}
\delta \lambda=\slashed{\partial}\Phi\,\epsilon-\frac{1}{2}\slashed{H}_3\sigma^3\,\epsilon
-\frac{1}{2}e^{\Phi} \Bigl(\slashed{F}_{3}\sigma^1+2i\slashed{F}_{1}\sigma^2\Bigr)\,\epsilon,
\ee
where $\sigma^i$ are the $2\times 2$ Pauli matrices which now act on the supersymmetry transformation parameter $\epsilon$ with the two Majorana-Weyl spinors of type IIB theory as components, \be\epsilon=\left(
                  \begin{array}{c}
                    \epsilon^1 \\
                    \epsilon^2 \\
                  \end{array}
                \right).\label{spinor10dee}\ee
The gamma matrices obey the anticommutation relation
\be
\{\Gamma^{\hat{M}},\,\Gamma^{\hat{N}}\}=2G^{\hat{M}\hat{N}}.
\ee
The slash is for contraction with the gamma matrices, and our definition for contracting
$p-q$ number of components of a $p$-form is
\be
(\slashed{\omega}_p)_{\hat{M}_{1}\cdots \hat{M}_q}= \frac{1}{(p-q)!}(\omega_p)_{\hat{M}_{1}\cdots \hat{M}_q\hat{M}_{q+1}\cdots \hat{M}_p}\Gamma^{[\hat{M}_{q+1}}\cdots \Gamma^{\hat{M}_p]},
\ee
where $\Gamma^{[\hat{M}_{q+1}}\cdots \Gamma^{\hat{M}_p]}\equiv\Gamma^{\hat{M}_{q+1}\cdots \hat{M}_p}$ is an antisymmetric product of the corresponding gamma matrices.
 $\nabla_{\hat{M}}$ is the covariant derivative given in terms of the spin connection $\omega^{\hat{M}\hat{N}}$ by
\be
\nabla_{\hat{M}}=(E^{\hat{M}}_{\,\,\,N})^{-1}\nabla_{N}=
(E^{\hat{M}}_{\,\,\, N})^{-1}\left(\partial_{N}+\frac{1}{4}\omega^{\hat{O}\hat{P}}_{\quad N}
\Gamma_{\hat{O}\hat{P}}\right),
\ee
where $\omega^{\hat{O}\hat{P}}_{\quad N}$ is defined through
\be
de^{\hat{M}}=-\frac{1}{2}\omega_{\hat{N}\hat{O}}^{\quad \hat{M}}\,
e^{\hat{N}}\wedge e^{\hat{O}}=-\omega^{\hat{M}}_{\,\,\,\hat{N}}\wedge e^{\hat{N}}
\ee
and
\be
\omega^{\hat{M}\hat{N}}=\omega_{\quad\hat{O}}^{\hat{M}\hat{N}}\,
dx^{\hat{O}},\quad
\omega_{\quad O}^{\hat{M}\hat{N}}=E_{\,\,\,O}^{\hat{P}}\,\omega^{\hat{M}\hat{N}}_
{\quad \hat{P}}.
\ee
The indices with  hats are raised and lowered using the flat metric $G_{\hat{M}\hat{N}}=G^{\hat{M}\hat{N}}$ of the coordinate one-forms.
For the metric given by (\ref{metric-3n}), we have the following nonvanishing components:
\ba
\omega^{\hat{\mu}\hat{4}}=\frac{1}{2}\frac{g_sN}{2\pi r_s}\sinh u\sqrt{\cosh^3u}\,e^{\hat{\mu}},\qquad \omega^{\hat{m}\hat{4}}=-\frac{1}{2}\frac{g_sN}{2\pi r_s}\sinh u \sqrt{\cosh^3u}\,e^{\hat{m}},
\ea
and
\ba
\omega_{\quad \mu}^{\hat{\mu}\hat{4}}=\frac{1}{2}\frac{g_sN}{2\pi r_s}\sinh u\cosh^2u,\qquad \omega_{\quad m}^{\hat{m}\hat{4}}=-\frac{1}{2}\frac{g_sN}{2\pi }\sinh u\cosh u,
\ea
with $m\ne 4$.

The killing spinor is the solution for $\epsilon$ of $\delta\lambda=0$ and $\delta \psi_{\hat{M}}=0$ in (\ref{st-psi}) and (\ref{st-lam}). With the fluxes in  (\ref{F5sol-1bbn}) and the spin connection above, the variation of the dilatino $\delta \lambda=0$ gives
\ba
\frac{1}{2}\cosh u\left(\Gamma^{\hat{4}\hat{6}\hat{7}}+\Gamma^{\hat{4}\hat{8}\hat{9}}\right)\sigma^3\epsilon+
\frac{1}{2}\sinh u\left(\Gamma^{\hat{5}\hat{6}\hat{7}}+\Gamma^{\hat{5}\hat{8}\hat{9}}\right)\sigma^1\epsilon+
i \Gamma^{\hat{5}}\sigma^2\epsilon=0.\label{lamvar-1}
\ea
The variation of the gravitino $\delta \psi_{\hat{M}}=0$ gives, taking $\epsilon$ to be a function of only the radial coordinate $\rho$,
\ba
&\sinh 2u \, \Gamma^{\hat{4}}\epsilon-\sinh u \left(\Gamma^{\hat{5}\hat{6}\hat{7}}+\Gamma^{\hat{5}\hat{8}\hat{9}}\right)\sigma^1\epsilon-
i \Gamma^{\hat{5}}\sigma^2\epsilon&\nn\\ &-\frac{i}{2} \cosh 2u
\left(\Gamma^{\hat{5}\hat{6}\hat{7}\hat{8}\hat{9}}+
\Gamma^{\hat{0}\hat{1}\hat{2}\hat{3}\hat{4}}\right)\sigma^2\epsilon=0&\label{psimuvar-1}
\ea
for $\hat{M}=\hat{\mu}$ and $\hat{M}=\hat{5}$,
\ba
&\sinh 2u  \, \Gamma^{\hat{4}}\epsilon+
2\cosh u \, \Gamma^{\hat{4}\hat{6}\hat{7}}\sigma^3\epsilon
-\sinh u \left(\Gamma^{\hat{5}\hat{6}\hat{7}}-\Gamma^{\hat{5}\hat{8}\hat{9}}\right)\sigma^1\epsilon+
i \Gamma^{\hat{5}}\sigma^2\epsilon&\nn\\ &-\frac{i}{2} \cosh 2u
\left(\Gamma^{\hat{5}\hat{6}\hat{7}\hat{8}\hat{9}}+
\Gamma^{\hat{0}\hat{1}\hat{2}\hat{3}\hat{4}}\right)\sigma^2\epsilon=0&\label{psi67var-1}
\ea
for $\hat{M}=\hat{6}$ and $\hat{M}=\hat{7}$,
\ba
&\sinh 2u  \, \Gamma^{\hat{4}}\epsilon+
2\cosh u \, \Gamma^{\hat{4}\hat{8}\hat{9}}\sigma^3\epsilon
+\sinh u \left(\Gamma^{\hat{5}\hat{6}\hat{7}}-\Gamma^{\hat{5}\hat{8}\hat{9}}\right)\sigma^1\epsilon+
i \Gamma^{\hat{5}}\sigma^2\epsilon&\nn\\ &-\frac{i}{2} \cosh 2u
\left(\Gamma^{\hat{5}\hat{6}\hat{7}\hat{8}\hat{9}}+
\Gamma^{\hat{0}\hat{1}\hat{2}\hat{3}\hat{4}}\right)\sigma^2\epsilon=0&\label{psi89var-1}
\ea
for $\hat{M}=\hat{8}$ and $\hat{M}=\hat{9}$, and
\ba
\Gamma^{\hat{4}}\partial_\rho\epsilon&=&\frac{1}{8}\frac{g_sN}{2\pi}\Bigl[2\cosh u \left(\Gamma^{\hat{4}\hat{6}\hat{7}}+\Gamma^{\hat{4}\hat{8}\hat{9}}\right)\sigma^3\epsilon+
\sinh u \left(\Gamma^{\hat{5}\hat{6}\hat{7}}+\Gamma^{\hat{5}\hat{8}\hat{9}}\right)\sigma^1\epsilon
\nn\\&&+i \Gamma^{\hat{5}}\sigma^2\epsilon +\frac{i}{2} \cosh 2u
\left(\Gamma^{\hat{5}\hat{6}\hat{7}\hat{8}\hat{9}}+
\Gamma^{\hat{0}\hat{1}\hat{2}\hat{3}\hat{4}}\right)\sigma^2\epsilon\Bigr]=0\label{psi4var-1}
\ea
for $\hat{M}=\hat{4}$. Observe that the total of eleven equations from the variations of the gravitino and the dilatino has been reduced to five at this point.

Combining (\ref{lamvar-1}), (\ref{psimuvar-1}), and (\ref{psi67var-1}), we obtain
\be
\sinh u\left(\Gamma^{\hat{5}\hat{6}\hat{7}}-\Gamma^{\hat{5}\hat{8}\hat{9}}\right)\sigma^1\epsilon=
\cosh u\left(\Gamma^{\hat{4}\hat{6}\hat{7}}-\Gamma^{\hat{4}\hat{8}\hat{9}}\right)\sigma^3\epsilon
\ee
which imposes
\be
\Gamma^{\hat{6}\hat{7}}\epsilon=\Gamma^{\hat{8}\hat{9}}\epsilon.\label{proj1}
\ee
Furthermore, the spinor operator in the 10D spacetime is given by
$
\Gamma_{(10)}=\Gamma^{\hat{0}\hat{1}\hat{2}\hat{3}\hat{4}\hat{5}
\hat{6}\hat{7}\hat{8}\hat{9}}
$
and its eigenvalues determine the chirality of the spinors. In type IIB theory, we have two Majorana-Weyl spinors of the same chirality, and we take
$\Gamma_{(10)}\epsilon=-\epsilon$. This means that
\be
\Gamma^{\hat{0}\hat{1}\hat{2}\hat{3}\hat{4}}\epsilon
=\Gamma^{\hat{5}
\hat{6}\hat{7}\hat{8}\hat{9}}\epsilon=-\Gamma^{5}\epsilon,
\ee
where we have used (\ref{proj1}) to write the last equality.  Moreover, the spinor operator on the worldvolume of the wrapped $D7-$branes is $\Gamma_{(8)}=\Gamma^{\hat{0}\hat{1}\hat{2}\hat{3}\hat{6}\hat{7}\hat{8}\hat{9}}$, and taking $\Gamma_{(8)}\epsilon=- i\epsilon$ with $\Gamma_{(10)}\epsilon=-\epsilon$ and (\ref{proj1}) sets
\be
\Gamma^{\hat{4}\hat{5}}\epsilon=-i\epsilon.\label{epssol-3}
\ee

With (\ref{proj1}) and (\ref{lamvar-1}), (\ref{psimuvar-1}) - (\ref{psi89var-1}) become identical, and together with (\ref{lamvar-1}) and (\ref{psi4var-1}), the eleven equations from the variations of the gravitino and the dilatino reduce to the following three equations,
\be
\cosh u\,\Gamma^{\hat{4}\hat{6}\hat{7}}\sigma^3\epsilon+
\sinh u\,\Gamma^{\hat{5}\hat{6}\hat{7}}\sigma^1\epsilon+
i \Gamma^{\hat{5}}\sigma^2\epsilon=0,\label{lamvar-2}
\ee
\be
\sinh 2u \, \Gamma^{\hat{4}}\epsilon+\cosh u\,\Gamma^{\hat{4}\hat{6}\hat{7}}\sigma^3\epsilon-\sinh u\,\Gamma^{\hat{5}\hat{8}\hat{9}}\sigma^1\epsilon+i \cosh 2u
\,\Gamma^{\hat{5}}\sigma^2\epsilon=0,\label{psiallno4-1}
\ee
\be
\partial_\rho\epsilon=-\frac{1}{8}\frac{g_sN}{2\pi}\Bigl[
\sinh 2u  + 4i \cosh^2u \, \Gamma^{\hat{5}}\sigma^2
\Bigr]\epsilon=0.\label{psi4var-2}
\ee

Next, we solve (\ref{lamvar-2}) - (\ref{psi4var-2}). First, we rewrite (\ref{lamvar-2}) as
\be
\Gamma^{\hat{4}\hat{5}\hat{6}\hat{7}}\sigma^1\epsilon-e^{iu \Gamma^{\hat{4}\hat{5}}\sigma^2} \epsilon=0,\label{lamvar-3}
\ee
and its solution is
\be
\epsilon=e^{-\frac{i}{2}u \Gamma^{\hat{4}\hat{5}}\sigma^2}\chi,\label{epssol-2}
\ee
where $\chi$ is a function of only $\rho$ such that
\be
\Gamma^{\hat{4}\hat{5}\hat{6}\hat{7}}\sigma^1\chi=\chi.\label{epssol-2a}
\ee
Now, (\ref{epssol-2}) with (\ref{epssol-2a}) also solves (\ref{psiallno4-1}). One way to see that is to rewrite (\ref{psiallno4-1}) as
\be
\Gamma^{\hat{4}\hat{5}\hat{6}\hat{7}}\sigma^1\,e^{-i u \Gamma^{\hat{4}\hat{5}}\sigma^2}\epsilon-e^{2i u \Gamma^{\hat{4}\hat{5}}\sigma^2}\epsilon=0,
\ee
which is solved by (\ref{epssol-2}) with (\ref{epssol-2a}).

Noting that $\partial_\rho=\frac{g_sN}{2\pi}\cosh^2 u\,\partial_u$, following (\ref{hn-1}), (\ref{psi4var-2}) becomes
\be
\partial_u\,\epsilon=-\frac{1}{8}\sinh 2u \sech^2 u\, \epsilon- \frac{i}{2}\Gamma^{\hat{5}}\sigma^2\epsilon.\label{epssol-4}
\ee
Using (\ref{epssol-2}) in (\ref{epssol-4}), we have the equation for $\chi$,
\be
\partial_u \chi=-\frac{1}{8}\sinh 2u \sech^2 u\, \chi,
\ee
and its solution is
\be
\chi={(\sech u)^{\frac{1}{4}}}\,\chi_0,\label{chisol-1}
\ee
where $\chi_0$ is a constant spinor.

With (\ref{chisol-1}) in (\ref{epssol-2}), the killing spinor is given by
\be
\epsilon={(\sech u)^{\frac{1}{4}}}\,e^{-\frac{i}{2}u \Gamma^{\hat{4}\hat{5}}\sigma^2}\chi_0,\label{epssol-2f}
\ee
where the constant spinor $\chi_0$ satisfies the three constraints
\be
\Gamma^{\hat{6}\hat{7}\hat{8}\hat{9}}\chi_0=-\chi_0,\qquad
\Gamma^{\hat{4}\hat{5}\hat{6}\hat{7}}\sigma^1\chi_0=\chi_0,\qquad
\Gamma^{\hat{4}\hat{5}}\chi_0=-i\chi_0,
\ee
following (\ref{proj1}), (\ref{epssol-2a}), and (\ref{epssol-3}).
These three constraints reduce the total of 32 supersymmetries to $4$, preserving $\mathcal{N}=1$ supersymmetry in four dimensions.
The killing spinor  $\epsilon = \chi_0$ on the wrapped $D7-$branes at $u=0$, and it varies along the radial direction away from the $D7-$branes.

\section{\label{sec:concl}Conclusion}

The gauge/gravity duality \cite{Maldacena:1998re,Gubser:1998bc,Witten:1998qj} provides a tool for investigating a gauge theory in terms of gravity theory and a gravity theory in terms of gauge theory.
We have shown directly using the supersymmetry variations of the gravitino and the dilatino of type IIB theory that the supergravity solutions presented in  \cite{Hailu:2011pn} preserve $\mathcal{N}=1$ supersymmetry in four dimensions.

We would like to reemphasize that the background
produces a robust linear confinement of quarks as shown in \cite{Hailu:2011kp} and further discussed in \cite{Hailu:2012ut}. Mass spectrum of $0^{++}$ glueballs that agrees with available data from large $N$ non-supersymmetric lattice QCD was obtained in \cite{Hailu:2012jc}.  A cosmological model in which the model universe undergoes dynamical accelerated expansion followed by a smooth transition to decelerating expansion was constructed in \cite{Hailu:2012ut}. The strong interaction was used for stabilization of the model universe during Friedmann evolution. The potential energy due of interaction between the model universe and the background serves as a source of dark energy.  Knowledge of the killing spinor presented here is useful for looking into supersymmetric and non-supersymmetric embeddings of $D-$branes and adding matter in the background.

In \cite{Hailu:2011pn}, we discussed how the location of the $D7-$branes along one of the five angular directions could accommodate chiral symmetry and its breaking in pure $\mathcal{N}=1$ $SU(N)$ gauge theory. However, the background geometry of the gravity theory has additional symmetries that are not obvious symmetries of pure $\mathcal{N}=1$ $SU(N)$ gauge theory, such as translational symmetry along the remaining four angular directions wrapped by the $D7-$branes. Orbifolding of the extra space for reducing the symmetry was proposed and discussed in \cite{Hailu:2011pn}. It would be useful to investigate the symmetries and parameter space of the background further and their implication on the gauge theory. Adding $D-$branes would also help in reducing the symmetry in addition to introducing matter.

\section*{Acknowledgements}

We are very grateful to Juan Maldacena for comments on the manuscript and suggesting looking into the additional symmetries of the supergravity background.

\providecommand{\href}[2]{#2}\begingroup\raggedright\endgroup

\end{document}